\documentclass[11pt,twoside]{article}

\usepackage{asp2006}
\usepackage{epsf}
\usepackage{graphicx}
\usepackage{lscape}

\newcommand{\ltsim}{\mbox
{{\raisebox{-0.4ex}{$\stackrel{<}{{\scriptstyle\sim}}$}}}}

\begin{document}
\title{Radio-loud and radio-quiet quasars:  one population, different
  epochs of observation}   %%% Fill in title 
\author{Katherine M.\ Blundell}   %%% Fill in author names
\affil{Oxford University Astrophysics}    %%% Fill in author affiliations

\begin{abstract} %%% Abstract to run on from here.
  I bring together evidence for the {\em rapidity} with which quasars'
  radio synchrotron lobe emission fades and for the {\em
    intermittency} with which jet plasma is ejected from individual
  quasars and radio galaxies and affirm the picture presented by
  \citet{nipoti2005} that the radio-loudness of quasars is a function
  of the epoch at which they are observed.  I briefly illustrate this
  account with examples of successive episodes of jet activity where
  the axis along which jet plasma is launched appears to have
  precessed.  A new model for the weak core radio emission from
  radio-quiet quasars, that is not any kind of jet ejecta, is also
  briefly described.
\end{abstract}

\section{How rapidly do radio lobes fade?}
\label{sec:lobefade}
It is a remarkably powerful observation that the radio sky does not
appear to be populated by vast numbers of dead or nearly dead radio
sources.  This is remarkable because radio galaxies and quasars are
not thought to be older than a few, perhaps several, $10^8$ years.
Since the age of the Universe is now widely believed to be 13.7
billion years \citep{bennett2003}, and given that radio galaxies and
quasars are routinely discovered out to redshifts corresponding to
lookback times of 12 billion years \citep[e.g.\, ][]{cruz2007}, it
would be reasonable to expect evidence of the cadavers of a good many
radio galaxies and quasars.  However, only a very few bona fide dead
radio galaxies are known \citep[e.g.\ ][]{cordey1987} although
careful, deep studies are being rewarded with a few more examples
\citep{parma2007}.

\subsection{Do we actually see a different picture at long radio
  wavelengths?} 

It is remarkably unusual that any radio galaxy or quasar has a
different morphology at low frequency ($< 100$s MHz) from its high
frequency (GHz) morphology.  It is worth noting that the challenges of
using interferometers to image extended emission at GHz frequencies
can disguise just how {\em unusual} is synchrotron plasma that emits
only at very low frequencies and not at all at GHz frequencies.  This
hints that low-frequency synchrotron emission fades (nearly) as
rapidly as high-frequency synchrotron emission and that synchrotron
cooling is not the dominant energy-loss mechanism for the synchrotron
plasma that radio lobes are composed of.  In fact, evidence from
matching the characteristics of complete samples of low-frequency
selected classical double radio sources in luminosity ($P$), linear
size ($D$), redshift ($z$) and spectral index ($\alpha$) with those of
simulated sources in these same characteristics led \citet{brw99} to
this realisation, which was investigated in more detail by
\citet{blundell2000}.

\begin{figure}[!ht]
\centering{
\includegraphics[width=10cm]{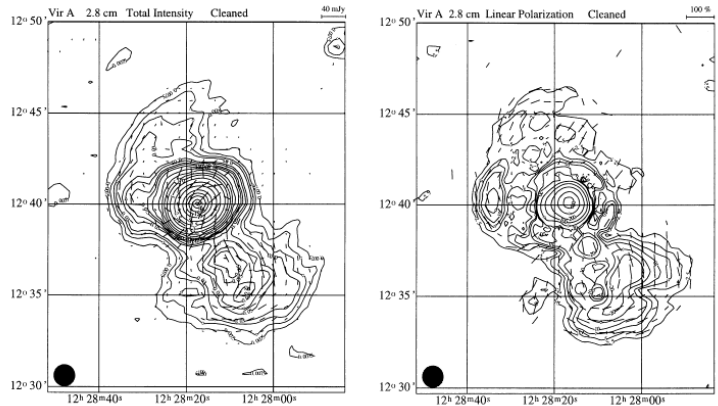}
\caption{The halo of M87 at 10\,GHz, whose imaging at this relatively
  high frequency by \citet{rottmann1996}, and also by
  \citet{andernach1979}, preceded the beautiful image at 330\,MHz by
  \citet{owen2000}.  }}
\end{figure}

The lobes of synchrotron emitting plasma associated with powerful
active galaxies such as classical double radio galaxies and quasars
are observed to be as much as several hundred kiloparsec in extent.
It was pointed out a number of decades ago \citep{Jen76} that {\em if}
synchrotron cooling played a part in determining the spectral shape of
extended lobes, then the lobes should be more extended at lower
frequencies.  This rarely appears to be the case!  For example,
observations of some 3C radio galaxies at 151\,MHz and 1.4\,GHz
\citep{Lea89} show that the lobe lengths at these different
frequencies are the same.  Fig.\,3 of \citet{bpk00} shows that the
images of 3C\,219 at 74\,MHz and at 1.5\,GHz are more remarkable for
their similarities than for their differences.  Just as in the cases
of 3C\,98 and 3C\,390.3 \citep{blundell2002} there is no evidence of
any extended emission at low frequency which is not already seen at
GHz frequencies; this appears to be the case for all the classical
doubles imaged to date.  If synchrotron cooling plays a very
significant role in determining the spectral shapes of lobes, then we
might expect to see {\sl lobes that extend further at low frequency
than at high frequency}.  There is no evidence for e.g.\ backflow
perpendicular to the source at 74\,MHz which is not detected at GHz
frequencies.  Observations suggest that the Lorentz factor particles
responsible for the 74\,MHz emission are entirely co-spatial with
those responsible for the 1.4\,GHz emission.  This is a first piece of
evidence suggesting that synchrotron particles of all energies
permeate the lobe magnetic field in the same way, despite the fact
that the high-$\gamma$ particles have shorter radiative lifetimes than
the lower energy ones.

\subsection{What do spectral index gradients really tell us?}

A second piece of evidence that synchrotron particles of all energies
permeate the lobe magnetic field in the same way may come from
observations of the way that spectral indices change along these lobes
\citep[e.g.\ ][]{Win80,Mye85,Ale87}: the general trend observed is
that the lobe spectra are flatter in the outermost regions near the
hotspot and steeper in the regions nearer the core.  Often the
observed change in spectral index with distance, or the spectral
gradient, is steady and systematic.

The traditional interpretation of spectral gradients goes as follows:
the radiating electrons nearer the core were dumped by the hotspot
much earlier in the past than the radiating electrons near the hotspot
now, and so the former will have undergone greater synchrotron cooling
compared with the latter.  A radiating population whose energy
distribution is initially a power-law, which suffered only synchrotron
losses, would result in a `break' in this power-law.  This break
frequency moves to lower frequencies as more time elapses
\citep{Kar62,Pac70} predicting steeper measured spectral indices for
the older emission near the centre of the source.  The location of
this break has been said to relate to the time elapsed since the
radiating particles were accelerated, in the so-called {\em spectral
  ageing} method \citep{Ale87,Mye85}, however, there are considerable
problems with this paradigm which have been explored
\citep{blundell2000} and summarised \citep{Blu01} elsewhere.  One such
problem is the inconsistency of this interpretation with the
observation that spectral gradients are observed well below the break
frequencies in some lobes.  For example, the images of Cygnus A
between 74\,MHz and 330\,MHz \citep{Kas96} shows a clear spectral
gradient at frequencies well below the fitted break frequencies
\citep{Car91}.  This behaviour would not be observed if the spectral
shapes at these frequencies were power-laws.  This observation is more
consistent with the assumption that there is a magnetic field gradient
along the lobe which `illuminates' different parts of a curved
spectrum (in Lorentz factor $\gamma$) at a given observing frequency
\citep{Blu01}.

The fast transport model, however, can potentially explain the
observations rather better: a gradient in magnetic field along the
lobe together with the same curved energy electron spectrum throughout
the lobe will result in a spectral gradient being observed at all
frequencies.  Indeed, analysis of multi-frequency images of Cygnus\,A
\citep{Rud94} show no evidence for any variation of the curved
$N(\gamma)$ spectrum across different regions of the lobe.  Another
remarkable result of this analysis is that the {\em bright} filaments
in the lobes of Cygnus\,A have {\em flatter} spectra than the
surrounding lobe material and also the same spectrum in Lorentz factor
$\gamma$ as elsewhere in the lobes, consistent with the idea that in
high $B$-field regions such as filaments, the flatter part of the
$\gamma$-spectrum is `illuminated' while in the lower $B$-field
regions higher-$\gamma$ particles are obviously required to give the
radiation at the particular $\nu_{\rm obs}$ and thus exhibit a steeper
spectrum.

There is no trivial identity which connects the radiative lifetimes of
the syncrotron-emiting particles with the age of the source, as
described by \citet{blundell2000}.

Consistent with the short radiative lifetimes of the synchrotron
particles is the observation that there are hardly any known dead
radio galaxies which have been observed.  There are barely a handful
of objects (such as \citet{cordey1987,parma2007}) that have extended
lobes while lacking any evidence for current on-going particle
acceleration (e.g.\ in hotspots or cores) even in the low-frequency
sky.  This confirms the radiative lifetimes of synchrotron particles
in the lobes being significantly shorter than the radiant lifetimes of
the radio galaxies themselves.

\subsection{How does synchrotron plasma age?}

If it is true that for classical double (FRII) radio sources we see
the same picture at long wavelengths (corresponding to $\sim
100\,$MHz) as we do at GHz frequencies, then this is a further
indication that energy losses of the synchrotron particles in the
lobes are {\em energy-independent} (such as would arise from adiabatic
losses) rather than {\em energy-dependent} (such as would arise from
synchrotron or inverse Compton cooling).   

It is important to realise that if the dominant energy losses are
energy-independent then searches at long wavelengths will {\em not}
reveal examples of relic activity in the Universe.   It is only if
the dominant loss mechanisms are energy-dependent that the
low-frequency Universe will look significantly different from the GHz
Universe we already know.

\section{Evidence for intermittency in quasars and radio galaxies}

\begin{figure}[!ht]
\centering{
\includegraphics[width=10cm]{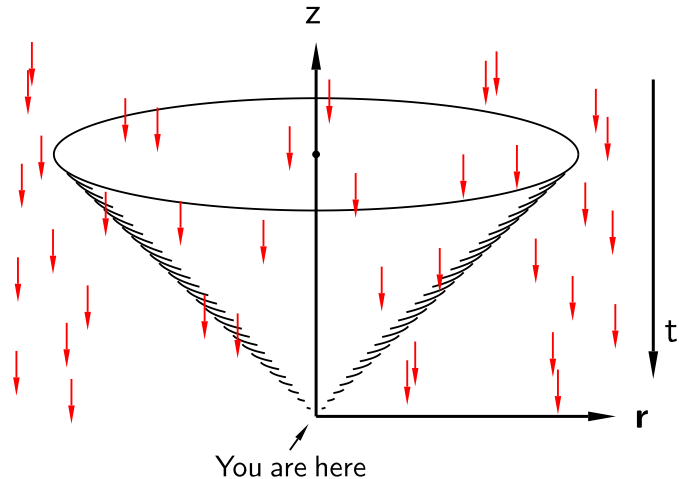}
\caption{ A schematic illustration of our light-cone. Each arrow
  represents the timeline of a radio source, whose lifetime is short
  compared to the Hubble time. Only those arrows which intercept our
  light-cone are those which we can observe. The point in a radio
  galaxy's lifetime when it is intercepted by our light-cone is of
  course random; thus, whether we see a quasar to be radio-loud or
  radio-quiet is random, with the relative numbers of radio-loud and
  radio-quiet quasars being determined by the duty-cycle of sustained
  jet ejection in quasars.  }}
\end{figure}

To glean evidence of intermittency in the jet activity of quasars is
more challenging to obtain than evidence of intermittency in the jet
activity of microquasars.  This is of course because microquasars
evolve on human-friendly timescales of hours, days and weeks whereas
the analogous phenomena in quasars take $10^5$ -- $10^7$ years, vastly
longer than human timescales.  

What might be the manifestation of episodic jet activity in quasars?
If the radio synchrotron plasma we observe from jets/lobes in quasars
arises from relativistic particles with high Lorentz factors (e.g.\
$10^4$ and above) then episodic jet activity would be evinced by the
detection of relic examples of such plasma, characterized by lower
Lorentz factors, nearer $10^3$ for example.  Particles with Lorentz
factors of $10^3$ are special because they inverse-Compton scatter
photons that comprise the peak of the Cosmic Microwave Background
(CMB) radiation to keV X-ray photons that are fairly easy to detect
with Chandra and XMM.  Searches for relic jet activity (manifested as
dead radio lobes) revealed by X-ray observations for such upscattered
emission have so far proved to be rather fruitful and are discussed
below.

\section{Episodic jet activity in quasars:  ``double-double'' examples}

A classical illustration of episodic activity in radio galaxies and
quasars are the so-called ``double-double'' radio galaxies.  There are
beautiful studies of such examples by \citet{schoemakers2000},
\citet{saripalli2002}, \citet{saripalli2003}, \citet{saikia2006} and
\citet{jamrozy2007}.

It seems that nature knows how to regenerate jet activity both along
fairly similar jet axes (especially in the case of double-doubles) and
also along rather different axes where some precession of the axis has
taken place as discussed in the next section.  Further investigation
might reward us with a more detailed understanding of how supermassive
black holes are fed from their environments.

\section{Episodic jet activity in quasars:  ``new direction'' examples}

\begin{figure}[!ht]
\centering{
\includegraphics[width=\textwidth]{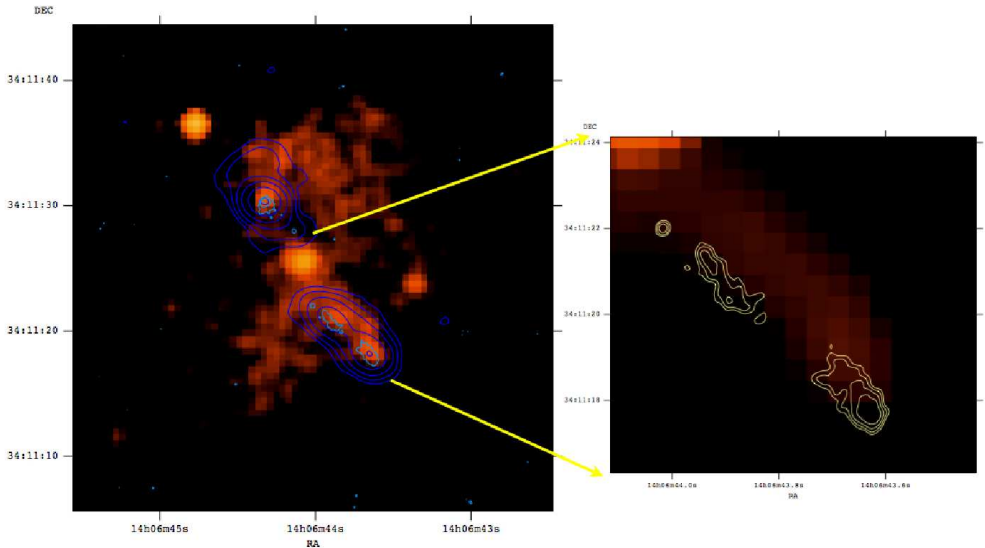}
\caption{\label{fig:3c294} The powerful radio galaxy 3C294 ($z =
  1.786$) observed in X-rays (greyscale) and radio (contours).
  Further figures are analysed by \citet{erlund2006}.  }}
\end{figure}

\citet{erlund2006} found that there is considerable extended X-ray
emission associated with the powerful radio galaxy 3C294; this is
reproduced in Figure\,\ref{fig:3c294}.  The bulk of the X-ray emission
shows extension along an axis differently oriented from the radio axis
by $\sim 50$\,degrees, and slightly longer than the length of the
radio axis.  Erlund et al suggest that the offset between these two
axes arises because of the precession of the axis along which jet
plasma is ejected.  The fact that the superposed radio and X-ray
observations reveal {\em intermittent} jet activity comes from the
fact that for the two discrete directions indicated, one of them is
traced by freshly accelerated high-energy radio-synchrotron emitting
particles while the other direction appears to be consistent with
inverse-Compton scattered CMB (ICCMB) photons (having a power-law
$\Gamma = 2.1 \pm 0.1$).  [Note that attributing the extended X-ray
emission to X-ray synchrotron would require a very spatially extended
acceleration mechansim for which no evidence has previously been seen
or invoked: the radiative lifetime of X-ray synchrotron emitting
particles is very short (\ltsim\ $100$ years).]  Consistent with the
notion that the jet axis is precessing is the observation revealed in
the zoom-in to Figure\,\ref{fig:3c294}, that at high resolution the
radio axis is offset from the nearby fine scale X-ray emission.

\begin{figure}[!ht]
\centering{
\includegraphics[width=\textwidth]{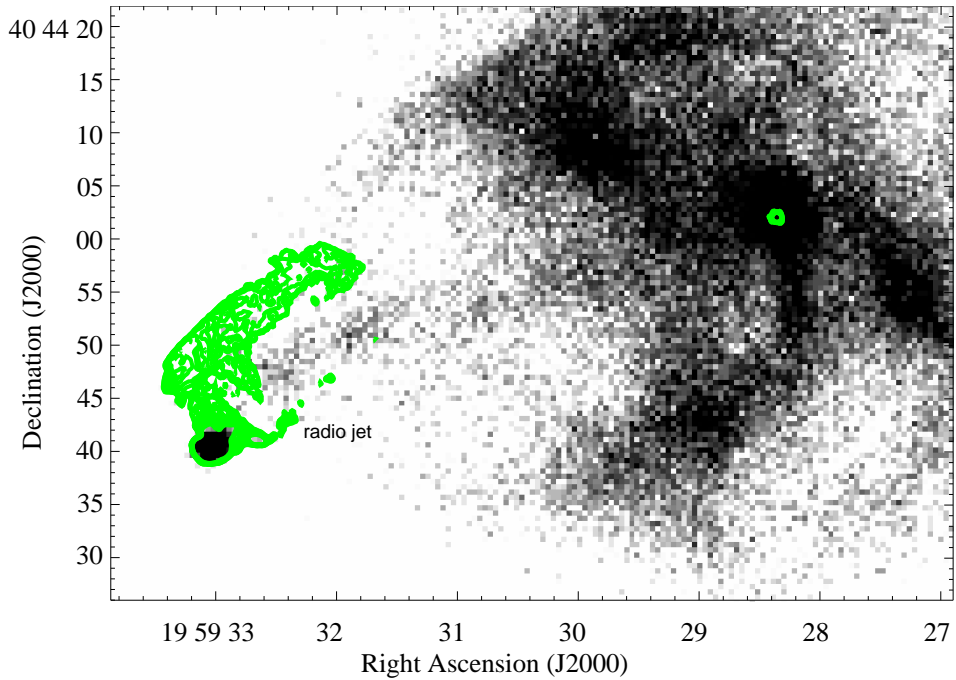}
\caption{\label{fig:cygarelic} Images of Cygnus\,A in X-rays
  (greyscale) and 5-GHz radio (contours) of the part of it mainly
  pointing away from Earth.  Further figures and analysis of its
  previous jet activity and variation in the jet axis are in
  \citet{steenbrugge2008a} and \citet{steenbrugge2008b}.  }}
\end{figure}

More recently, \citet{steenbrugge2008a} and \citet{steenbrugge2008b}
have analysed superimposed X-ray and radio observations of the
prototypical FRII radio galaxy Cygnus\,A and identified, from the
co-addition of all relevant observations from the Chandra archive, a
linear feature in X-rays.  This feature easily satisfies the
\citet{bridle1984} criteria for classification as a jet, and its
energy spectrum is power-law (rather than thermal).  For these and
other reasons, \citet{steenbrugge2008b} interpret this linear feature
as a relic jet, albeit along a somewhat different direction from that
delineated by synchrotron radio emission.  Figure\,\ref{fig:cygarelic}
depicts as contours the radio synchrotron emission (the radio
counter-jet is the most southerly of the radio emission, other than
the East hotspot) while the relic X-ray counterjet is seen just to the
north of this.  \citet{steenbrugge2008a} present an analysis of the
variations in the directions of the jets in terms of a precession of
the launch axis of the jet ejecta.

Both 3C294 and Cygnus\,A show evidence of a different launch direction
of jet ejecta revealed by the presence of low-energy (Lorentz factor
$10^3$) particles giving rise to ICCMB.  There are other radio sources
that may show evidence of a different jet axis, with both axes being
revealed at radio wavelengths.  These are the so-called winged or
X-shaped radio sources and examples of this class are 3C403 and
3C223.1; these have been studied by \citet{dennettthorpe2002} who
favoured a fast, symmetric realignment of the jet axis on a timescale
of a few Myr in these objects (although see \citet{kraft2005} who
reach a different view).

\section{Precession versus Scheuer's dentist's drill: more than a
  semantic distinction?}

\begin{figure}[!ht]
\centering{
\includegraphics[width=\textwidth]{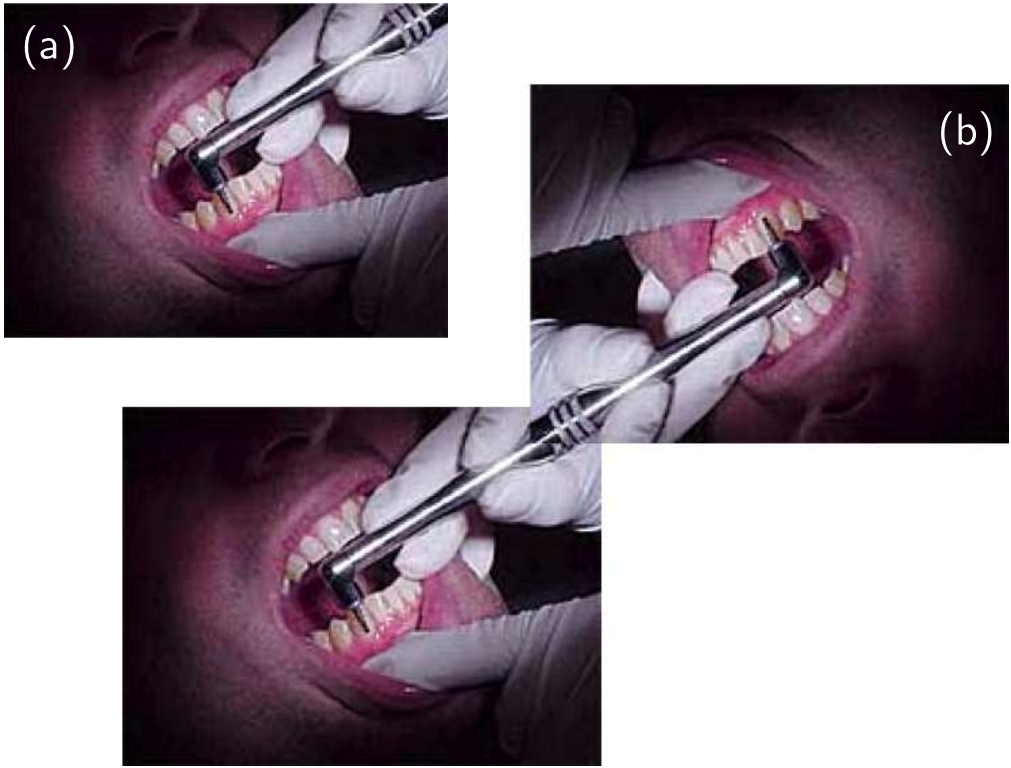}
\caption{\label{fig:dentistsdrill}(a) The dentists' drill model in its
  originally conceived one-sided form, devised by Scheuer (1982).  (b)
  A less-plausible, 2-sided version used by an imaginary dentist.  }}
\end{figure}

Precession, of course, in its most general sense includes any change
of the instantaneous spin axis.  Generally defined precession includes
the entire spectrum of spin-axis variations from polar wandering and
nutation to Earth's Chandler wobble.  All of these examples of
precession are fundamentally two-sided (applying to the emerging jet
and counterjet equally and simultaneously in the rest-frame of the
nucleus).  Temporal variation in precession parameters, as long as
two-sided and instigated at the jet launch point by angular momentum
changes (e.g.\ caused by variation in the fuelling), are properly
described as precession and are distinct from Scheuer's (1982) Dentist's
Drill phenomenon.  That phenomenon is a response of a jet, on {\em one
  side} of a source, to local conditions (for example, buoyancy
effects corresponding to local motions or inhomogeneities);
Figure\,\ref{fig:dentistsdrill} is a reminder that dentists' drills on
Earth are one-sided.

If deviations in jet-direction are neglected, and dismissed merely as
a dental drill meandering, there exists the possibility that
significant two-sided (symmetric) reorientations are overlooked.
Symmetric direction change might arise from varying angular momentum
in the matter being accreted by the supermassive black hole.

I remark that (symmetric, two-sided) precession of jet axes can of
course occur without this precession being as steady or periodic as in
the remarkably persistent case of the microquasar SS433
\citep{hjellming1981}.

\subsection{Two modes of energy loss: (I) the flaring mode (jets)}

\citet{blundell2000,blundell1999} have presented evidence (from
physical arguments and from the observed near-absence of any dead
radio galaxies) that lobe emission must disappear on relatively rapid
timescales, as mentioned in Section\,\ref{sec:lobefade}  In the light
of the relatively rapid disappearance of radio lobes when jets are
switched off, it is interesting to consider that the residual
lobe-less quasars may be conveniently identified with radio-quiet
quasars.

\citet{nipoti2005} suggested a parallel between radio-loudness in
quasars and the flaring mode (i.e.\ jet-ejecting mode) in
microquasars, and hence an association of radio-quiet quasars with
non-flaring states of microquasars.  This is in agreement with the
suggested association of radio-quiet AGN with `high/soft' states
\citep{Mac03}, but it conflicts with the association \citep{Fal04} of
FR\,I radio sources with low-hard states (small accretion rate and
steady radio emission).

\subsection{Two modes of energy loss: (II) the coupled mode (cores)}

\citet{nipoti2005} suggested that the radio emission in quasars that
is associated with the `coupled' mode identified for microquasars, is
confined to core (i.e.\ nuclear) emission from quasars, and is most
readily detectable at GHz frequencies.  Indeed, this radio core
emission is observed from many quasars classified as radio-quiet
\citep{Blu98}.  They suggested that the `flaring' mode leads to the
formation of the large-scale ($> {\rm kpc}$) jets that are the
hallmark of radio-loud quasars, be they FR\,Is or FR\,IIs.

Recently, \citet{blundell2007} advanced a new, physical model for the
radio emission from the cores of radio-quiet quasars (the `coupled'
mode in Nipoti et al's picture) that is significantly different from
models related to the notion of a cosmically conspiritorial sequence of
synchrotron self-absorbed jet knots originally advanced by
\citet{cotton1980}.  Radio emission from radio-quiet quasars is very
weak and, if present, is confined to the nucleus or core region; it
has been revealed by milli-arcsecond scale imaging techniques to arise
from regions no larger than a few cubic parsec in extent
\citep{Blu98}.

Blundell \& Kuncic's (2007) model posits that optically thin
bremsstrahlung from a slow, dense disc wind can make a significant
contribution to the observed levels of radio luminosity arising from
the unresolved cores of radio-quiet quasars.  This model was inspired
by observations of resolved disc wind emission observed directly via
milli-arcsecond radio imaging of SS433
\citep{blundell2001,paragi2002}.  If this thermal disc wind model
turns out to be widely applicable for radio-quiet quasars, it will
explain the long standing conundrum that radio-quiet quasars are
strongly accreting yet lack the very obvious means of mass-loss and
angular momentum-loss via directional jets: on the contrary, mass-loss
via on-going disc winds accompanies persistent disc accretion.

\acknowledgements %%% Text of acknowledgements runs on after this command.
I wish to record my thanks to the conference organisers for their
support and for organising such an enjoyable meeting and to the Royal
Society for a University Research Fellowship.  I warmly thank my
collaborators James Binney, Peter Duffy, Mary Erlund, Andy Fabian,
Zdenka Kuncic, Carlo Nipoti, Steve Rawlings and Katrien Steenbrugge
and especially Ian Heywood \& Robert Laing for a careful reading of
this manuscript.


\begin{thebibliography}{}

\bibitem[Alexander \& Leahy(1987)]{Ale87}
Alexander, P.\ \& Leahy, J.P., 1987, \mnras, 225,1

\bibitem[Andernach et al.(1979)]{andernach1979} Andernach, H., Baker, 
J.~R., von Kap-Herr, A., \& Wielebinski, R.\ 1979, \aap, 74, 93 

\bibitem[Bennett et al.(2003)]{bennett2003} Bennett, C.~L., et al.\ 
2003, \apjs, 148, 1 

\bibitem[Blundell \& Beasley(1998)]{Blu98} 
Blundell, K.~M. \& Beasley, A.~J.\ 1998, \mnras, 299, 165 

\bibitem[Blundell \& Kuncic(2007)]{blundell2007} Blundell, K.~M., \& 
Kuncic, Z.\ 2007, \apjl, 668, L103 

\bibitem[Blundell et al.(2000a)]{bpk00} Blundell, K., Kassim, 
N., \& Perley, R.\ 2000, arXiv:astro-ph/0004005 

\bibitem[Blundell et al.(2001)]{blundell2001} Blundell, K.~M., 
Mioduszewski, A.~J., Muxlow, T.~W.~B., Podsiadlowski, P., \& Rupen, M.~P.\ 
2001, \apjl, 562, L79 

\bibitem[Blundell \& Rawlings(1999)]{blundell1999} Blundell, K.~M., 
\& Rawlings, S.\ 1999, \nat, 399, 330 

\bibitem[Blundell \& Rawlings(2000)]{blundell2000} Blundell, K.~M., 
\& Rawlings, S.\ 2000, \aj, 119, 1111 

\bibitem[Blundell \& Rawlings(2001)]{Blu01}
Blundell K.M.\ \& Rawlings S., 2001 in {\it Particles \& Fields in Radio
  Galaxies}, ASP 250, 363; astro-ph/0209372

\bibitem[Blundell et al.(1999)]{brw99} Blundell, K.~M., 
Rawlings, S., \& Willott, C.~J.\ 1999, \aj, 117, 677 

\bibitem[Blundell et al.(2002)]{blundell2002} Blundell, K.~M., 
Rawlings, S., Willott, C.~J., Kassim, N.~E., \& Perley, R.\ 2002, New 
Astronomy Review, 46, 75 

\bibitem[Bridle \& Perley(1984)]{bridle1984} Bridle, A.~H., \& 
Perley, R.~A.\ 1984, \araa, 22, 319 

\bibitem[Carilli et al(1991)]{Car91}
Carilli C.L., Perley R.A., Dreher J.W., \& Leahy J.P., 1991, 
\apj, 383, 554

\bibitem[Cotton et al.(1980)]{cotton1980} Cotton, W.~D., et al.\ 
1980, \apjl, 238, L123 

\bibitem[Cruz et al.(2007)]{cruz2007} Cruz, M.~J., Jarvis, 
M.~J., Rawlings, S., \& Blundell, K.~M.\ 2007, \mnras, 375, 1349 

\bibitem[Cordey(1987)]{cordey1987} Cordey, R.~A.\ 1987, \mnras, 
227, 695 

\bibitem[Dennett-Thorpe et al.(2002)]{dennettthorpe2002} Dennett-Thorpe, 
J., Scheuer, P.~A.~G., Laing, R.~A., Bridle, A.~H., Pooley, G.~G., \& 
Reich, W.\ 2002, \mnras, 330, 609 

\bibitem[Erlund et al.(2006)]{erlund2006} Erlund, M.~C., Fabian, 
A.~C., Blundell, K.~M., Celotti, A., \& Crawford, C.~S.\ 2006, \mnras, 371, 
29 

\bibitem[Falcke, K\"ording \& Markoff(2004)]{Fal04} 
Falcke H., K\"ording E., Markoff S., 2004, A\&A, 414, 895

\bibitem[Fender et al.(2004b)]{Fen04b} 
Fender, R., Wu, K., Johnston, H., Tzioumis, T., Jonker, P., Spencer,
R., \& van der Klis, M.\  2004b, \nat, 427, 222 

\bibitem[Fomalont et al.(2001)]{fomalont2001} Fomalont, E.~B., 
Geldzahler, B.~J., \& Bradshaw, C.~F.\ 2001, \apj, 558, 283 

\bibitem[Hjellming \& Johnston(1981)]{hjellming1981} Hjellming, 
R.~M., \& Johnston, K.~J.\ 1981, \apjl, 246, L141 

\bibitem[Hjellming \& Rupen(1995)]{hjellming1995} Hjellming, R.~M., 
\& Rupen, M.~P.\ 1995, \nat, 375, 464 

\bibitem[Jamrozy et al.(2007)]{jamrozy2007} Jamrozy, M., Konar, C., 
Saikia, D.~J., Stawarz, {\L}., Mack, K.-H., \& Siemiginowska, A.\ 2007, 
\mnras, 378, 581 

\bibitem[Jenkins \& Scheuer(1976)]{Jen76}
Jenkins, C.J., \& Scheuer, P.A.G.\ 1976, \mnras, 174, 327 

\bibitem[Kardashev(1962)]{Kar62}
Kardashev N.S., 1962, \sovast, 6, 317

\bibitem[Kassim et al.(1996)]{Kas96}
Kassim N.E., Perley R.A., Carilli C.L., Harris D.E.\ \&
Erickson W.C., 1996, in {\em `Cygnus A -- Study of
a radio galaxy'}, p154, eds C.L.\ Carilli \& D.E.\ Harris

\bibitem[Kraft et al.(2005)]{kraft2005} Kraft, R.~P., Hardcastle, 
M.~J., Worrall, D.~M., \& Murray, S.~S.\ 2005, \apj, 622, 149 

\bibitem[Leahy et al.(1989)]{Lea89}
Leahy J.P., Muxlow T.W.B., \& Stephens P.W., 1989,
\mnras, 239, 401

\bibitem[Maccarone, Gallo \& Fender(2003)]{Mac03}
Maccarone T.J., Gallo E., Fender R., 2003, MNRAS, 345, L19

\bibitem[Miller-Jones et al.(2004)]{millerjones2004} Miller-Jones, 
J.~C.~A., Blundell, K.~M., Rupen, M.~P., Mioduszewski, A.~J., Duffy, P., \& 
Beasley, A.~J.\ 2004, \apj, 600, 368 

\bibitem[Mioduszewski et al.(2001)]{mioduszewski2001} Mioduszewski, 
A.~J., Rupen, M.~P., Hjellming, R.~M., Pooley, G.~G., \& Waltman, E.~B.\ 
2001, \apj, 553, 766 

\bibitem[Mirabel \& Rodr{\' i}guez(1994)]{mirabel1994} 
Mirabel, I.~F.~\& Rodr{\' i}guez, L.~F.\ 1994, \nat, 371, 46 

\bibitem[Mirabel \& Rodr{\' i}guez(1999)]{mirabel1999} 
Mirabel, I.~F., \& Rodr{\' i}guez, L.~F.\ 1999, ARAA, 37, 409 

\bibitem[Myers \& Spangler(1985)]{Mye85}
Myers S.T.\ \& Spangler S.R., 1985, \apj, 291, 52

\bibitem[\protect\citeauthoryear{Nipoti, Blundell \& Binney}{Nipoti et
    al.}{2005}]{nipoti2005} Nipoti C, Blundell KM \& Binney JJ, 2005,
  MNRAS, 361, 633

\bibitem[Owen et al.(2000)]{owen2000}
Owen, F.~N., Eilek, J.~A., 
\& Kassim, N.~E.\ 2000, \apj, 543, 611 

\bibitem[Pacholczyk(1970)]{Pac70}
Pacholczyk A.G., 1970, `Radio Astrophysics', pub.\
W. H. Freeman \& Co 

\bibitem[Paragi et al.(2002)]{paragi2002} Paragi, Z., Fejes, I., 
Vermeulen, R.~C., Schilizzi, R.~T., Spencer, R.~E., \& Stirling, A.~M.\ 
2002, Proceedings of the 6th EVN Symposium, 263 

\bibitem[Parma et al.(2007)]{parma2007} Parma, P., Murgia, M., de 
Ruiter, H.~R., Fanti, R., Mack, K.-H., \& Govoni, F.\ 2007, \aap, 470, 875 

\bibitem[Rottmann et al.(1996)]{rottmann1996} Rottmann, H., Mack, 
K.-H., Klein, U., \& Wielebinski, R.\ 1996, \aap, 309, L19 

\bibitem[Rudnick et al.(1994)]{Rud94}
Rudnick L., Katz-Stone D.M.\ \& Anderson M.C., 1994,
\apjs, 90, 955

\bibitem[Saikia et al.(2006)]{saikia2006} Saikia, D.~J., Konar, 
C., \& Kulkarni, V.~K.\ 2006, \mnras, 366, 1391 

\bibitem[Saripalli et al.(2002)]{saripalli2002} Saripalli, L., 
Subrahmanyan, R., \& Udaya Shankar, N.\ 2002, \apj, 565, 256 

\bibitem[Saripalli et al.(2003)]{saripalli2003} Saripalli, L., 
Subrahmanyan, R., \& Udaya Shankar, N.\ 2003, \apj, 590, 181 

\bibitem[Schoenmakers et al.(2000)]{schoemakers2000} Schoenmakers, 
A.~P., de Bruyn, A.~G., R{\"o}ttgering, H.~J.~A., van der Laan, H., \& 
Kaiser, C.~R.\ 2000, \mnras, 315, 371 

\bibitem[Scheuer(1982)]{scheuer1982} Scheuer, P.~A.~G.\ 1982, 
Extragalactic Radio Sources, 97, 163 

\bibitem[Steenbrugge \& Blundell(2008)]{steenbrugge2008a}
Steenbrugge K.C.\ \& Blundell K.M., 2008, \mnras, in press

\bibitem[Steenbrugge et al.(2008)]{steenbrugge2008b}
Steenbrugge K.C., Blundell K.M.\ \& Duffy P., 2008, \mnras, submitted

\bibitem[Winter et al.(1980)]{Win80}
Winter A.J.B. et al., 1980, \mnras, 192, 931

\end{thebibliography}
\end{document}